\documentclass[12pt]{article}
\usepackage{amsmath}
\usepackage{amsfonts}
\usepackage{amssymb}
\usepackage{latexsym}
\usepackage{graphicx}
\usepackage{enumerate}
\usepackage[shortlabels]{enumitem}
\usepackage{mathtools}
\usepackage{hyperref}
\usepackage{color}
\usepackage{xcolor}
\usepackage{fullpage}
\usepackage{tikz}
\usepackage{pdflscape}

\usepackage{amsmath, amssymb,latexsym}
\usepackage{color}
\usepackage{xcolor}
\usepackage{fullpage}
\usepackage{soul}
\usepackage{float}
\usepackage{comment}

\def\R{\mathbb{R}}

\newtheorem{theorem}{Theorem}[section]
\newtheorem{proposition}[theorem]{Proposition}

\newtheorem{lemma}[theorem]{Lemma}
\newtheorem{definition}[theorem]{Definition}

\newtheorem{remark}[theorem]{{\sc Remark}}

\definecolor{brilliantrose}{rgb}{1.0, 0.33, 0.64}
\definecolor{myviolet}{rgb}{0.21, 0.0, 0.85}
\definecolor{amethyst}{rgb}{0.6, 0.4, 0.8}
\definecolor{carrotorange}{rgb}{0.93, 0.57, 0.13}

\title{Efficient computation of Katz centrality for very dense networks via {negative parameter Katz}}
\author{Vanni Noferini\thanks{Corresponding author. Aalto University, Department of Mathematics and Systems Analysis, P.O. Box 11100, FI-00076, Aalto, Finland. Supported by an Academy of Finland grant (Suomen Akatemian p\"{a}\"{a}t\"{o}s 331240).}
\and 
Ryan Wood\thanks{Aalto University, Department of Mathematics and Systems Analysis, P.O. Box 11100, FI-00076, Aalto, Finland. Supported by an Academy of Finland grant (Suomen Akatemian p\"{a}\"{a}t\"{o}s 331240).}
}

\begin{document}
\maketitle
\begin{abstract} 
Katz centrality (and its limiting case, eigenvector centrality) is a frequently used tool to measure the importance of a node in a network, and to rank the nodes accordingly. One reason for its popularity is that Katz centrality can be computed very efficiently when the network is sparse, i.e., having only $O(n)$ edges between its $n$ nodes. While sparsity is common in practice, in some applications one faces the opposite situation of a very dense network, where only $O(n)$ potential edges are missing with respect to a complete graph. We explain why and how, even for very dense networks, it is possible to efficiently compute the ranking stemming from Katz centrality for unweighted graphs, possibly directed and possibly with loops, by working on the complement graph. Our approach also provides an interpretation, regardless of sparsity, of ``Katz centrality with negative parameter" as usual Katz centrality on the complement graph. For weighted graphs, we provide instead an approximation method that is based on removing sufficiently many edges from the network (or from its complement), and we give sufficient conditions for this approximation to provide the correct ranking. We include numerical experiments to illustrate the advantages of the proposed approach.
\end{abstract}

\textbf{Keywords:} Katz centrality, dense network, complement graph, negative parameter, eigenvector centrality


\section{Introduction}

Given a finite graph, a \emph{centrality measure} is a non-negative function defined on the nodes. In practice, this corresponds to a non-negative vector $v$ whose $i$-th component, $v_i$, is the centrality of node $i$. In network analysis, centrality measures have the goal of quantifying the importance of each node within the network \cite{EH}. One approach towards the definition and computation of centrality measures is based on the enumeration of traversals across the network. So-called walk-based centrality measures have gained popularity in network analysis, also due to the fact that walks on a network can be efficiently enumerated by linear-algebraic techniques \cite{Benzi,estrada}. A classical example of walk-based centrality measures is Katz centrality \cite{Katz,newman}, whose limit corresponds to eigenvector centrality \cite{Bonacich,newman}. 

Recall that a network is \emph{sparse} if it contains far fewer edges than the complete graph with the same number of nodes \cite{newman}; more concretely, one may define sparsity as the property, for a network with $n$ nodes, of having at most $O(n)$ edges out of the $O(n^2)$ possible connections. One advantage of Katz or eigenvector centrality is that methods are known for their efficient computation for potentially very large, but sparse, networks \cite{Benzi}; these algorithms rely on the mathematical connection between graph theory and matrix theory \cite{EH}, and on the efficient implementation of modern numerical linear algebra libraries such as LAPACK \cite{lapack99}. As a vast majority of the networks studied in practice are sparse \cite{newman}, and a significant fraction is very large in size \cite{Benzi,newman}, their low computational complexity explains in part the popularity of Katz and eigenvector centrality. However, while sparsity is a property of most real-life networks \cite{newman}, exceptions are not unheard of. Indeed, it is possible to encounter applications \cite{ANV,science,Dunne,newman,PDMA,RS,plos1} that are modelled by networks that are the very opposite of sparse: They are \emph{very dense} in the sense of having most (say, all but $O(n)$ or less) edges present among all the possible edges in the network. 

 The present article deals with the problem of efficiently computing the ranking of the nodes stemming from Katz centrality when the network is very dense. After reviewing some background material in Section \ref{sec:background}, we obtain new results in Section \ref{sec:newstuff}. {In Section \ref{sec:realworld} our approach is tested numerically on some graphs arising in finance.} Our main message is that an exact efficient computation is possible for very dense unweighted graphs; this is stated formally in Theorems \ref{thm:trick1} and \ref{thm:trick2}. We also argue that for certain non-sparse weighted graphs this task is also possible, via the statement of Theorems \ref{thm:trick3} and \ref{thm:trick4} possibly followed by a certain approximation. Theorem \ref{thm:sufficient1} provides a sufficient condition for this approximation to provide the same ranking of the nodes as Katz centrality. While Katz centrality is usually only considered for positive values of the Katz parameter $t$, a byproduct of our analysis is to also give meaning to Katz centrality for \emph{negative} $t$; indeed, our main results show that this induces a centrality measure equivalent to Katz on the complement graph.

\section{Background material}\label{sec:background} Let us start by recalling some basic definitions in graph theory \cite{estrada}.

\begin{definition}
    A \emph{finite graph} is a pair of sets $G=(V,E)$ where the set $V$ has finite cardinality and $E \subseteq V \times V$. If, for all $i \in V$, $(i,i) \not \in E$ then we say that $G$ is \emph{without loops}, otherwise it is \emph{with loops}.

A \emph{walk} of length $k$ in a graph is a sequence (possibly with repetitions) $i_0,i_1,\dots,i_k 
 \subset  V$ such that $(i_{r-1},i_{r}) \in E$ for all $r=1,\dots,k$.
\end{definition}

In this note, we will need the notion of complement graph. We give two possible definitions, depending on whether we restrict ourselves to graphs without loops.

\begin{definition}
    If $G=(V,E)$ is a finite graph, then its \emph{complement graph} is $G^c=(V,E^c)$ where $E^c$ is the complement of $E$ in $V \times V$.

    If $G = (V,E)$ is a finite graph without loops, let $P=(V \times V) \setminus \{ (i,i) | i \in V  \}$. Then the \emph{complement graph without loops} of $G$ is $G^{cw}=(V,E^{cw})$ where $E^{cw}$ is the complement of $E$ in $P$.
\end{definition}

The set of unweighted finite graphs is isomorphic with the set of $\{0,1\}$ matrices in the following way.

\begin{definition}
    Let $G=(V,E)$ be finite a graph with $V=\{1,\dots,n \}=:[n]$. Its \emph{adjacency matrix}, denoted by $A$, is the $n \times n$ matrix such that $A_{ij}=1$ if $(i,j) \in E$ and $A_{ij}=0$ otherwise. The \emph{spectral radius} of $A$, denoted $\rho(A)$, is the modulus of the largest (in modulus) eigenvalue of $A$.
\end{definition}
By the Perron-Frobenius theorem \cite{Friedland}, since $A$ is non-negative, we also know that the spectral radius $\rho(A)$ itself is in fact necessarily an eigenvalue of $A$. It is an elementary exercise to prove that, if $A$ is the adjacency matrix of $G$, then $(A^k)_{ij}$ is the number of walks of length $k$ from $i$ to $j$ in $G$. 
Based on this observation, as well as the intuition that the more walks begin from a given node, the more central that node should be to the network, L. Katz proposed in \cite{Katz} to compute the centrality of node $i$ as the weighted sum of all the walks starting from $i$, where the weight is $t^k$ for some $t \in ]0,1[$; in other words, Katz suggested to exponentially downweight walks according to their length. In formulae, \emph{Katz centrality} of node $i$ is thus
\begin{equation}\label{eq:katz}
    v_i = \sum_{j=1}^n \sum_{k=0}^\infty t^k (A^k)_{ij}.
\end{equation}
Assume now that the \emph{Katz parameter} $t$ satisfies $0 < t < \rho(A)^{-1}$, which is necessary for convergence, and define $e \in \R^n$ as the vector whose entries are all $1$. Rewriting \eqref{eq:katz} in vector form, we  obtain the formula
\begin{equation}\label{eq:katzvec}
    v = \left( \sum_{k=0}^\infty t^k A^k \right) e = (I - t A)^{-1}e.
\end{equation}
These manipulations show that Katz centrality can be computed by solving a linear system with coefficient matrix $I-tA$ and right hand side $e$. It is known \cite{DDHK} that linear system solving has at most the same theoretical asymptotic complexity as matrix multiplication (and possibly less); nevertheless, in practice, it is often performed in $O(n^3)$ operations with traditional methods such as Gaussian elimination. However, for networks arising from concrete applications, $A$ (and thus $I-tA)$ is typically sparse having $O(n)$ nonzero elements. This is exploited numerically by numerical linear algebra libraries, leading to a much more competitive complexity \cite{Benzi}. 

While sparsity is a feature of most real-world networks \cite{newman}, there are nonetheless exceptions. Indeed, there exist networks of practical interest that not only are not sparse, but in fact are \emph{very dense}, by which we mean that the number of non-edges, i.e., edges that the graph \emph{lacks}, is $O(n)$ or less. Examples include recommendation systems \cite{plos1}, certain online social networks \cite{plos1}, food webs in ecosystems \cite{Dunne}, some models for human brain connectivity \cite{science,RS}, or networks used in finance for investment portfolio optimization \cite{ANV,PDMA}. Let us briefly discuss this latter example (in a slightly simplified manner: see \cite{ANV,PDMA} for more details), which is of particular interest as motivation for the present paper given the fact that it directly discusses centrality measures on dense networks. The authors of \cite{PDMA} proposed, and successfully tested, a sophisticated technique to select stocks for investment purposes; further developments and tests were presented in \cite{ANV}. Two steps within this method are based on network theory: (1) Forming a graph based on the correlation of a large set of stocks, and (2) Ranking the nodes of this network (that is, the stocks available on a certain exchange market) by, say, their Katz centrality. One possible method to construct the graph is to place an edge between stock $i$ and $j$ if and only if the absolute value of the correlation between the time series of the prices of $i$ and $j$ on the stock market for the previous year is greater than a certain threshold. As it turns out that many pairs of stocks are quite strongly correlated, such ``financial networks" are frequently very dense. Therefore, one cannot directly exploit the sparsity of the associated adjacency matrices for the computation of Katz centrality. Fortunately, one can nonetheless efficiently rank the nodes according to Katz centrality on such network. As we shall see below, this is achievable by computing Katz centrality on the (sparse) complement graph, after selecting a \emph{negative} value of the Katz parameter.

Finally, we recall \cite{Bonacich} that the ranking produced by Katz centrality in the limit when $t \rightarrow \rho(A)^{-1}$ is the same as that produced by \emph{eigenvector centrality} \cite{BenziKlymko}. The latter is maybe even more popular than Katz thanks to its parameter-free nature, and it is defined as the centrality measure corresponding to a Perron eigenvector of $A$, i.e., a non-negative vector $v$ such that $Av=\rho(A) v$. Beyond Katz, another variant of eigenvector centrality is used, for example, in the celebrated PageRank algorithm \cite{pagerank}. We point out that the results in this paper can be applied to eigenvector centrality as well, by taking the {limit $t \rightarrow \rho(A)^{-1}$} of the Katz parameter.

\section{Computing Katz and eigenvector centrality of very dense networks via the complement graph}\label{sec:newstuff}

\subsection{Graphs with loops}

Let us start with the case where loops are allowed. We consider a graph $G$ with adjacency matrix $A$, and its complement graph $G^c$ that by definition has adjacency matrix $B:=ee^T-A$. It turns out that Katz centrality on $G$ can be indirectly, and possibly much more efficiently, computed via Katz centrality on $G^c$ but with a \emph{negative} value of the Katz parameter.

\begin{theorem}\label{thm:trick1}
    Let $G$ be a finite graph with adjacency matrix $A$, and let $G^c$ be its complement graph. Then, for all $0 < t < \rho(A)^{-1}$, Katz centrality on $G$ with parameter $t$ yields the same ranking as Katz centrality on $G^c$ with parameter $-t$.
\end{theorem}
\begin{proof}
    Noting that $A=ee^T-B$, by the Sherman-Morrison formula \cite{SM50}
    \[  (I - t A)^{-1} = (I + t B - t e e^T)^{-1} = (I+tB)^{-1} + t \frac{(I+tB)^{-1}ee^T(I+tB)^{-1}}{1-t e^T (I+tB)^{-1}e}.   \]
    If we now define $\gamma=\gamma(t,B):=e^T(I+tB)^{-1}e$, this in turn implies
    \begin{equation}\label{eq:trick1}
        (I-tA)^{-1} e = \left( 1+ \frac{t \gamma}{1 - t \gamma} \right) (I+ t B)^{-1} e = \frac{1}{1-t \gamma} (I + t B)^{-1}e.    
    \end{equation} 
The left hand side of \eqref{eq:trick1} is the vector of Katz centralities of $G$ with parameter $t$. We claim that $t \gamma < 1$ and that $(I+tB)^{-1}e$ is a centrality measure, i.e., a non-negative vector. Hence, up to the scalar factor $(1-t \gamma)^{-1}$, the right hand side of \eqref{eq:trick1} is ``Katz centrality with negative parameter $-t$" of $G^c$.

It remains to prove the claim. Indeed, while $(I-tA)^{-1}=\sum_{k=0}^\infty (tA)^k$ for $|t|<\rho(A)^{-1}$ has a non-negative Taylor expansion around $0$, $(I+tB)^{-1} = \sum_{k=0}^\infty (-tB)^k$ (which is valid for small enough $t$) does not, and therefore the claim is not fully obvious a priori. For a proof, note that from \eqref{eq:trick1} we get that \[ \frac{e}{1-t \gamma} = (I+tB)(I-tA)^{-1}e;\]
but, since $|t|<\rho(A)^{-1}$, the vector $(I+tB)(I-tA)^{-1}e$ is the product of non-negative matrices and vectors, and hence must be non-negative. It follows that $t \gamma < 1$, and that $(I+tB)^{-1} e = (1-t\gamma)(I-tA)^{-1} e$ is non-negative.
\end{proof}

It is convenient to give a few remarks. First, a practical advantage stemming from Theorem \ref{thm:trick1} is that $(I+tB)^{-1} e$ can be computed efficiently when $A$ has only $O(n)$ zero elements, as in this case $I+tB$ has $O(n)$ nonzero elements.

Second, the statement and proof of Theorem \ref{thm:trick1} show that, although Katz centrality is usually only considered for positive values of its parameter $t$, it still yields a centrality measure for (small enough, where the upper bound on the allowed values depends, with inverse proportionality, on the spectral radius of the \emph{complement} of the considered graph) negative values of the parameter. Combinatorially, Katz centrality with a negative parameter $t$ corresponds to the difference of a weighted sum of even-length walks minus a weighted sum of odd-length walks. This can be seen by the identity

\[ (I + t B)^{-1} = (I - t B) (I - t^2 B^2)^{-1} =  \sum_{k \ \mathrm{even}} t^k B^k  - \sum_{k \ \mathrm{odd}} t^k B^k, \]
which holds algebraically over the field of fraction of the ring of formal power series, and analytically for all sufficiently small $t$.  
Of course, Theorem \ref{thm:trick1} also gives an alternative interpretation of yielding a centrality which is equivalent to classical, positive-parameter, Katz on the complement graph. { Alternative interpretations of Katz centrality with a negative parameter are discussed in \cite{NEG}, where it is suggested that a negative $t$ can be used for modelling competitive situations.}

Finally, note that \eqref{eq:trick1} can also be interpreted as a valid algebraic identity over the field $\R(t)$, that is, it holds as an identity of rational functions in the variable $t$. This implies that whenever $1/t$ is an  eigenvalue of $A$, but not an eigenvalue of $-B$, then necessarily $1=t \gamma(t)$. In particular, $\gamma(\rho(A)^{-1},B)=\rho(A)$ in this case. We give below an alternative proof of this fact assuming that $\rho(B)<\rho(A)$. 

\begin{lemma}\label{lem:alternative}
     In the same notation as above, let $\gamma(t,B):=e^T (I+tB)^{-1} e$. If $\rho(B) < \rho(A)$, then $\gamma(\rho(A)^{-1},B)=\rho(A)$.
\end{lemma}
\begin{proof}
    Let $v$ be a non-negative Perron eigenvector of $A$ \cite{Friedland}, normalized to have unit $1$-norm, i.e., $A v = \rho(A) v$ and $e^T v =1$. Then, we see that $B v = e - \rho(A) v$. Hence, for all $0 \leq t  < 1/\rho(B)$
    \[ (I+tB)^{-1} e = \sum_{k \ \mathrm{even}} t^k (B^k \rho(A) v + B^{k+1} v) - \sum_{k \ \mathrm{odd}} t^k (B^k \rho(A) v + B^{k+1} v) =\]
    \[  = \rho(A) v + \sum_{k=1}^\infty (-B)^k v (t^k \rho(A) - t^{k-1})  .\]
    Substituting $t=\rho(A)^{-1} < \rho(B)^{-1}$ yields \[ \gamma\left(\frac{1}{\rho(A)},B\right) =  e^T \left(I+\frac{B}{\rho(A)}\right)^{-1} e = \rho(A) e^T v = \rho(A).\]
    
\end{proof}

{The assumption $\rho(B)<\rho(A)$, needed for Lemma \ref{lem:alternative}, is very reasonable in the scenario that motivates this paper, where $B$ has $O(n)$ nonzero element. We make a more precise statement in Proposition \ref{prop:rhoAmorethanrhoB}.

\begin{proposition}\label{prop:rhoAmorethanrhoB}
    Let $B=ee^T-A$ be an $n \times n$ adjacency matrix of an unweighted graph with loops, and suppose that $B$ has strictly fewer than $\frac{n^2}{4}$ nonzero elements. Then $\rho(B) < \rho(A)$.
\end{proposition}
\begin{proof}
 Denote by $\beta^2$ the number of nonzero elements in $B$, so that $2 \beta < n$. Observe that $\rho(B) \leq \|B\|_2 \leq \beta$. On the other hand, using also the Bauer-Fike theorem \cite{BF60},
    \[  \rho(ee^T)=n \geq \rho(A) \geq \rho(ee^T) - \| B \|_2 \geq n - \beta. \]
    It follows that
    $\rho(A) - \rho(B) \geq n - 2 \beta > 0$.
\end{proof}

}

To conclude the present subsection, we note in passing that eigenvector centrality can also be computed by means of the complement graph. If the power method is used and the Perron eigenvector $v$ is normalized so that $0 \leq v$, $e^Tv=1$, for example, and as long as one starts from a normalized $v_0$ s.t. $e^T v_0=1$, then the iteration $v_k \leftarrow v_{k+1}= \frac{A v_k}{e^T A v_k}$ is equivalent to the potentially much cheaper iteration 
\begin{equation}\label{eq:power}
    v_k \leftarrow v_{k+1} = \frac{e - B v_k}{n-e^T B v_k}.
\end{equation} 

Alternatively, if $-\rho(A)$ is not an eigenvalue of $B$, the ranking of eigenvector centrality can also be computed from \eqref{eq:trick1}, as $(I+ \frac{B}{\rho(A)})^{-1}e$.

\subsection{Graph without loops}

We now turn our attention to graphs without loops. If $G$ has adjacency matrix $A$, then its complement graph without loops $G^{cw}$ has adjacency matrix $B:=ee^T-I-A$. Even in this case, one can compute Katz centrality on $G$ by computing a Katz centrality on $G^{cw}$ with a negative Katz parameter.

\begin{theorem}\label{thm:trick2}
    Let $G$ be a finite graph without loops with adjacency matrix $A$, and let $G^{cw}$ be its complement graph without loops. Then, for all $0< t < 1 / \rho(A)$, Katz centrality on $G$ with parameter $t$ yields the same ranking as Katz centrality on $G^{cw}$ with parameter $-\frac{t}{1+t}$.
\end{theorem}

\begin{proof}
    Since $A+B=ee^T-I$, we have
\[ (I -t A)^{-1} = (I(1+t)+tB-t ee^T)^{-1} = \frac{1}{1+t} \left( I + \frac{t}{1+t} B - \frac{t}{1+t} ee^T \right)^{-1}.\]
Define now $\chi=\chi(t,B):=e^T \left( I + \frac{t}{1+t} B \right)^{-1}e$. Then, again using the Sherman-Morrison formula, we get
\begin{equation}\label{eq:trick2}
    (I-tA)^{-1} e = \frac{1}{t+1} \frac{1}{1-t \chi} \left( I + \frac{t}{t+1} B \right)^{-1}e.
\end{equation}
Similarly to Theorem \ref{thm:trick1}, we now argue that $t \chi < 1$, because \eqref{eq:trick2} implies that 
\[ 0 \leq \left(I + \frac{t}{t+1} B \right) (I-t A)^{-1} e = \frac{e}{(t+1)(1-t \chi)};\]
in turn, this also implies that $(I+\frac{t}{t+1} B)^{-1} e$ is a non-negative vector, and hence, a centrality measure.
\end{proof}

Analogous remarks as in the previous subsection hold:
\begin{itemize}
    \item By our previous discussion, the formulation $(I+\frac{t}{t+1} B)^{-1} e$ is much more efficient to compute than $(I-tA)^{-1} e$ when $A$ is very dense;
    \item Katz centrality for negative values of the Katz parameter has the combinatorial interpretation of summing (with weight) even-length walks and subtracting (with weight) odd-length walks;
    \item Again, \eqref{eq:trick2} also holds as a valid algebraic identity over $\R(t)$, implying that $1=t \chi(t)$ whenever $1/t$ is an eigenvalue of $A$. In particular, $\chi(\rho(A)^{-1},B)=\rho(A)$;
    \item Eigenvector centrality can also be computed by working on the complement graph without loops. In this case, as long as one starts with a normalized $v_0$ s.t. $e^T v_0=1$, the power method iteration becomes
\begin{equation}\label{eq:powernoloops}
    v_k \leftarrow  v_{k+1} =   \frac{ e - v_k - B v_k}{n-1-e^T B v_k}.
\end{equation}  

  \item An alternative approach to compute the ranking of eigenvector centrality is via \eqref{eq:trick2}, as $\left(I + \frac{B}{\rho(A)+1}\right)^{-1}e$, provided that $-\rho(A)-1$ is not an eigenvalue of $B$.
\end{itemize}

\subsection{Weighted graphs}\label{sec:weightedgraphs}

In applications, often graphs are weighted, that is, each edge $(i,j) \in E$ is associated with a positive real weight $\omega(i,j)>0$. In this case, the adjacency matrix is defined as $A_{ij}=\omega(i,j)$ if $(i,j)$ is an edge and $A_{ij}=0$ otherwise. Katz centrality is still defined as \eqref{eq:katz}, and in this case one weights a walk of length $k$ by $t^k$ times the product of the weights of the edges that are travelled in the walk; {for more details see, e.g., \cite{Ryan} and the references therein}. Using this more general definition, Katz centrality can still be computed as \eqref{eq:katzvec} for all $0<t<\rho(A)^{-1}$. Without loss of generality, since scaling both $A$ and $t$ in inverse proportion yields the same centrality, we may assume that $\max A_{ij}=1$. In this setting, it still makes perfect sense to define the complement graph as the graph whose adjacency matrix is $B=ee^T-I-A$ in the case of graph without loops. In the case of graph with loops, again one could opt for $B=ee^T-A$; however, it is interesting to note that the graph whose adjacency matrix $B=eu^T -A$, where $(u)_j = \max_i (\omega(i,j))$, is at least as sparse {as} $ee^T-A$. In fact, note that $u \leq e$ elementwise and that, for any vector $z$ such that $u \leq z \leq e$, $e z^T-A$ is an adjacency matrix; on the other hand, it is clear that choosing $z=u$ minimizes the number of edges in the corresponding graph.

The statements and proofs of Theorems \ref{thm:trick1} and \ref{thm:trick2} do \emph{not} require that $A$ and $B$ have elements in $\{0,1\}$ (although they do require that $0 \leq A,B \leq 1$ elementwise; but it is straightforward to rescale if this is not the case). Similarly, the rank one matrix $ee^T$ can be replaced by $eu^T$. Consequently, we can give a version of both Theorems for weighted graphs below, as Theorems \ref{thm:trick3} and \ref{thm:trick4} respectively. To avoid a cumbersome statement, we formally identify a non-edge in a weighted graph with ``an edge having weight zero".

We state Theorem \ref{thm:trick3} assuming, for simplicity of exposition, that $\Omega=1$; as previously mentioned, this is no loss of generality because obviously one can always rescale the weights without changing the rankings of Katz centrality.

\begin{theorem}\label{thm:trick3}
    Let $G$ be a weighted finite graph with loops, such that $u_j = \max_i (\omega((i,j)))$. Assume without loss of generality that $\Omega=\max_j u_j=1$, and let $G^c$ be the complement graph of $G$, defined so the weight of $(i,j)$ in $G^c$ is $$u_j-\omega(i,j) = \max_i (\omega(i,j)) - \omega(i,j) \geq 0$$

    if and only if the weight of $(i,j)$ in $G$ is $\omega(i,j)$. Then, for all $0 < t < \rho(A)^{-1}$, Katz centrality on $G$ with parameter $t$ yields the same ranking as Katz centrality on $G^c$ with parameter $-\gamma t$, where $\gamma = \dfrac{1}{1+u^T(I-tA)^{-1}e}$ is guaranteed to be positive.
\end{theorem}

Theorem \ref{thm:trick4} is slightly simpler to state for a generic value of $\Omega=\max_{i,j} \omega(i,j)$.
\begin{theorem}\label{thm:trick4}
    Let $G$ be a weighted finite graph without loops, such that the largest weight is $\Omega>0$. Let $G^{cw}$ be the complement graph without loops of $G$, defined so that, for all $i \neq j$, the weight of $(i,j)$ in $G^{cw}$ is $\Omega-\omega(i,j)$ if and only if the weight of $(i,j)$ in $G$ is $\omega(i,j)$. Then, for all $0 < t < \rho(A)^{-1}$, Katz centrality on $G$ with parameter $t$ yields the same ranking as Katz centrality on $G^c$ with parameter $-\frac{ t}{1 + \Omega t }$.
\end{theorem}

\begin{remark}
    The adjacency matrix of the complement graph is defined to be $B=eu^T-A$ when using Theorem \ref{thm:trick3} (and assuming $\Omega=1$, which can always be achieved by rescaling $A$ if necessary) and $B=\Omega ee^T-\Omega I - A$ when using Theorem \ref{thm:trick4}. In practice, a loopless graph could also seen as a graph with loops that happens to not have loops; hence, a choice can be made in order to minimize the number of edges in the complement. Note that, if $A$ has zero diagonal, then $\Omega ee^T-\Omega I -A$ has at least $n+1$ zero elements (but potentially more), whereas $eu^T-A$ has at least $n$ zero elements (but potentially more).

If the thresholding technique described below is to be used, then a choice can be made in order to maximize the number of edges having small (or zero) weight.
\end{remark}

Theorem \ref{thm:trick3} and Theorem \ref{thm:trick4} are interesting generalizations of Theorem \ref{thm:trick1} and Theorem \ref{thm:trick2}, but at first sight their direct applicability seems to be more limited. The problem is that, for weighted graphs, unless all but $O(n)$ edges have exactly the same maximal weight $\Omega$ (if using Theorem \ref{thm:trick4}) or column maximal weight $u_i$ (if using Theorem \ref{thm:trick3}), this technique does not result in a sparse adjacency matrix $B$ for the complement graph. Hence, while the theoretical results still hold, the practical motivation of obtaining a cheaper computational method to compute the centrality may falter. Still, if many edges in $G$ have weight \emph{close} (but not exactly equal) to $\Omega$ (or $u_i)$, one may approximate $B$ with a sparser matrix $B_0$ by setting to $0$ all weights {that are less than} a certain threshold $\varepsilon$. Clearly, this results in a sparse network if only $O(n)$ edges in the complement graph have weights $> \varepsilon$. Of course, this technique could alternatively be applied directly also to the \emph{original} graph, when it has $O(n)$ edges with weights $> \varepsilon$; we omit the details as it is clear that in this second case one simply switches the roles of $A$ and $B$.

A crucial question is: The technique described above will likely produce approximated, but inexact, values of the Katz centrality vector, but when is the approximation error sufficiently small to not change the ranking of the nodes? Theorem \ref{thm:sufficient1} below gives a sufficient condition for the approximation to yield the desired node ranking, i.e., the same as Katz, in the case of weighted graph without loops. For simplicity of exposition, we assume again $\Omega=1$ in the statement and proof of Theorem \ref{thm:sufficient1}.

\begin{theorem}\label{thm:sufficient1}
Let $A,B=eu^T-A$ be defined as in Theorem \ref{thm:trick3} for a weighted finite graph with loops, with $\max_{i,j} A_{i,j}=1$, and hence $0 \leq A,B \leq 1$. Let $\varepsilon > 0$ and let $B_0$ be such that 
\[ (B_0)_{ij} = \begin{cases}
    B_{ij} \ &\mathrm{if} \ B_{ij} > \varepsilon;\\
    0 \ &\mathrm{if} \ B_{ij} \leq \varepsilon.
\end{cases} \]
       Define
    \[ v = (I + t B)^{-1}e, \quad v_0 = (I+t B_0)^{-1}e, \quad w = (I-t A)^{-1} e,  \quad x  = \frac{\min_{i \neq j} |w_i-w_j|}{\max_i w_i}, \]
    and 
    \[  c_j = \varepsilon^{-1} \max_i  (B-B_0)_{ij}.   \]
If \begin{equation}\label{eq:sufficient1}
t < \rho(A+\varepsilon ec^T)^{-1} \quad and \quad   \varepsilon < \frac{x}{t (c^T w) (1+x)},
\end{equation} 
then $v$ and $v_0$ yield the same ranking.
\end{theorem}

\begin{proof}
Observe that the statement reduces to a vacuously true implication if $x=0$. Hence, we restrict ourselves to $x>0$, and in particular we assume that every node has a distinct Katz centrality for the chosen value of $\alpha$, i.e., all the entries of the vector $w$ are distinct.

Suppose now that \eqref{eq:sufficient1} holds with $x>0$. Let $C:=\varepsilon^{-1} (B-B_0)$ so that elementwise $0 \leq C \leq e c^T$. By Theorem \ref{thm:trick3}, $v$ and $w$ yield the same ranking; for the same reason, $v_0$ and $w_0:=(I-tA-t \varepsilon C)^{-1} e$ also yield the same ranking. Thus, it suffices to prove that $w$ and $w_0$ yield the same ranking. To this goal, note that elementwise $0 \leq A \leq A + \varepsilon C \leq A + \varepsilon ec^T$. Hence,
\[  \sum_{k=0}^\infty t^k A^k \leq \sum_{k=0}^\infty t^k (A+\varepsilon C)^k \leq \sum_{k=0}^\infty t^k (A+\varepsilon ec^T)^k, \]
implying in turn that elementwise
\begin{equation}\label{eq:interesting}
    w \leq w_0 \leq (I - t A - t \varepsilon e c^T)^{-1} e = w + \varepsilon t \frac{(I-tA)^{-1}ec^T(I-tA)^{-1}e}{1-t \varepsilon c^T (I-tA)^{-1} e} = w \left(1 + \varepsilon t\frac{c^T w}{1-\varepsilon t c^T w} \right),
\end{equation}  
where we have used the Sherman-Morrison formula for the second last step. It is convenient to define now $y:=(c^T w) \varepsilon t$; note that, by assuming \eqref{eq:sufficient1}, $0 \leq y < \frac{x}{1+x} < 1$. Then, we can rewrite \eqref{eq:interesting} as 
\[  w \leq w_0 \leq w \left(1 + \frac{y}{1-y}\right).\]  Suppose now for a contradiction that $w$ and $w_0$ do not yield the same ranking. Then, there exists a pair $i \neq j$ such that $w_i > w_j$ but $(w_0)_i \leq (w_0)_j$. Therefore,
\[ 0 \leq \frac{(w_0)_j - (w_0)_i}{w_j} \leq \frac{w_j - w_i}{w_j} +   \frac{y}{1-y} \leq - x +  \frac{y}{1-y}.\]
But the function $y \mapsto f(y):=\frac{y}{1-y}$ is increasing in $[0,1[$, which together with $y < \frac{x}{1+x}$ implies the contradictory inequality $0<0$, because
\[ 0 \leq -x + f(y) < -x + f\left(\frac{x}{1+x}\right) = -x + x = 0.\]

\end{proof}

    Theorem \ref{thm:sufficient1} can be extended to the loopless case as follows (again assuming without loss of generality that $\Omega = 1$). 
\begin{theorem}\label{thm:sufficient2}
Let $A,B=e e^T-I-A$ be as above for a weighted finite graph with no loops, with $\max_{i,j} A_{i,j}=1$ and $0 \leq A,B \leq 1$. Let $\varepsilon > 0$ and let $B_0$ be such that 
\[ (B_0)_{ij} = \begin{cases}
    B_{ij} \ &\mathrm{if} \ B_{ij} > \varepsilon;\\
    0 \ &\mathrm{if} \ B_{ij} \leq \varepsilon.
\end{cases} \]
       Define
    \[ v = \left(I + \frac{t}{1+t} B\right)^{-1}e, \quad v_0 = \left(I+\frac{t}{1+t} B_0\right)^{-1}e, \quad w = (I-t A)^{-1} e,  \quad x  = \frac{\min_{i \neq j} |w_i-w_j|}{\max_i w_i},  \]
    and 
    \[  c_j = \varepsilon^{-1} \max_i  (B-B_0)_{ij}.   \]
If \begin{equation}\label{eq:sufficient2}
t < \rho(A+\varepsilon ec^T)^{-1} \quad and \quad   \varepsilon < \frac{x}{t (c^T w) (1+x)},
\end{equation} 
then $v$ and $v_0$ yield the same ranking.
\end{theorem}

\begin{proof}
	As for Theorem \ref{thm:sufficient1}, we may assume $x>0$ and define $C := \varepsilon^{-1}(B-B_0)$ and $w_0 := (I-tA - t\varepsilon C)^{-1}$. Note that

\begin{align*}
		w_0 &= (I - tA +tB_0 - tB)^{-1}e\\
  & = ((1+t)I+t B_0 - t ee^T)^{-1}e \\ 
		&= \frac{1}{1+t}\left(I - \frac{t}{1+t} (B_0-ee^T)\right)^{-1}e.
	\end{align*}
By Theorem~\ref{thm:trick4}, $v$ yields the same centrality as $w$; as do $w_0$ and $v_0$ by Theorem~\ref{thm:trick3}. Therefore it remains only to prove that $w$ and $w_0$ yield the same centrality. From this point on, the proof is analogous to the proof of Theorem~\ref{thm:sufficient1}. 
\end{proof}

\begin{remark}
    Note that, in \eqref{eq:sufficient1}, the right hand side is the product of $\frac{1}{t(c^T w)}$ and $\frac{x}{1+x}$. Unlike $w$ and $x$, $c$ is a (nonlinear) function of $\varepsilon$. Hence, while testing \eqref{eq:sufficient1} a posteriori is certainly possible, it may be difficult to use it a priori to determine a good value of $\varepsilon$.

One possibility is to note that, for all $\varepsilon$, $c \leq e$, implying $(e^T w)^{-1} \leq (c^T w)^{-1}$. Hence, it is possible to prove a variant of Theorem \ref{thm:sufficient1} that yields a different sufficient condition for exact recovery, namely,
\begin{equation}\label{eq:sufficient3}
    \varepsilon < \frac{x}{t(e^T w)(1+x)}.
\end{equation}
Clearly \eqref{eq:sufficient3} implies \eqref{eq:sufficient1}. In theory, \eqref{eq:sufficient3} can be much more demanding than \eqref{eq:sufficient1} --- but not necessarily so in practice; see Table \ref{tab2} for a comparison in the case of some randomly generated weighted graphs. On the other hand, \eqref{eq:sufficient3} has the advantage that its right hand side is independent of $\varepsilon$.

Analogous remarks hold for \eqref{eq:sufficient2}, which is also implied by \eqref{eq:sufficient3}; we omit the details.
\end{remark}

Ideally, one would like on one hand that $\varepsilon$ is sufficiently small to satisfy \eqref{eq:sufficient1} or \eqref{eq:sufficient2}, and on the other hand that it is sufficiently large to make $B_0$ sparse. This may not always be possible in practice, and furthermore it can be expensive to check \eqref{eq:sufficient1}, \eqref{eq:sufficient2}, or \eqref{eq:sufficient3} numerically in spite of their theoretical interest. On the more optimistic side, Theorem \ref{thm:sufficient1} and Theorem \ref{thm:sufficient2} give a sufficient, but potentially not necessary, condition for exact recovery of the ranking. Moreover, it may be sufficient for practical purposes to compute a ranking that, if not exactly equal, is at least reasonably close to the one coming from Katz. To test whether this can happen regardless of conditions \eqref{eq:sufficient1} or \eqref{eq:sufficient2}, in Table \ref{tab1} we report the results of an experimental test that suggests that, when $B_0$ is sparse with $O(n)$ edges, not only the corresponding approximation of Katz centrality is efficient to compute but it may also yield an excellent approximation of the correct rankings even when an exact recovery fails. {While the experiment of Table \ref{tab1} is artificial, in Section \ref{sec:realworld} we present more evidence, coming from tests of financial networks, that an inexact but efficient computation of Katz for weighted network can be an attractive compromise.}

    \begin{table}[h!]
        \centering
        \begin{tabular}{|c|c|c|c|}
        \hline
           size of $B$  & mean of $\tau$ & minimum of $\tau$ &  {average time to compute $v_0$}\\
            \hline
           300   &  0.9361 & 0.9251 & { 0.0036 s}\\
            \hline
           900  & 0.9354 & 0.9299 & { 0.0278 s}\\
            \hline
           1500  & 0.9356 & 0.9315 & { 0.0701 s}\\
           \hline
           3000  & 0.9355 & 0.9325 & { 0.5322 s}\\
           \hline
        \end{tabular}
        \caption{\emph{A positive matrix $D$ of size $3n \times 3n$ is generated as a realization of an i.i.d. random matrix whose entries have cdf $x^{1/n}$ on $[0,1]$; by construction, $D$ is a.s. full, but the expected number of its entries that are $\geq 0.1$ is $9 n^2 (1-(0.1)^{1/n}) \approx 9\log(10) n$ for large $n$. In practice, this is realized simply by the MATLAB command \texttt{D=rand(3*n).\^{}n}. We then set $A=ee^T-D$, $t=\frac12 \rho(A)^{-1}$, $B=eu^T-A$, and construct $B_0$ from $B$ by setting to $0$ every entry $\leq 0.1$. Then, we compute Kendall's correlation coefficient $\tau$ \cite{kendall} for the rankings provided by $v=(I+t B)^{-1}e$ and $v_0=(I+t B_0)^{-1}e$; recall that $-1\leq\tau\leq 1$ and that $\tau=1$ for equal rankings, $\tau=-1$ for opposite rankings, and $\mathbb{E}[\tau]=0$ for two independent non-constant random vectors. In the table, we report both the average and the minimum values of $\tau$ over $100$ instances of the experiment for each value of $n$. We remark that, while the computation of $v_0$ is efficient thanks to the sparsity of $B_0$ {(we report in the table the computational times with MATLAB R2019b on a MacBook Air laptop equipped with 1.6 GHz Intel Core i5)}, computing $v$ for a hundred times becomes prohibitively slow (on an ordinary laptop and without any parallelization) for values of $n$ significantly larger than in this experiment.}}
        \label{tab1}
    \end{table}

Table \ref{tab2} tests instead, assuming that exact recovery is necessary, where the compromise between sparsity of $B_0$ and applicability of Theorem \eqref{thm:sufficient1} lies. 

    \begin{table}[h!]
        \centering
        \begin{tabular}{|c|c|c|c|c|}
        \hline
        \multicolumn{5}{|c|}{$\varepsilon=(3n)^{-1}$}\\
        \hline
           size $n$ & sparsity & RHS of \eqref{eq:sufficient1} & RHS of \eqref{eq:sufficient3} & $\varepsilon$  \\
            \hline
           300   &  94.5 $\%$ & $2.2 \cdot 10^{-7}$ & $1.7 \cdot 10^{-7}$ & $3.3 \cdot 10^{-3}$\\
            \hline
           900  & 97.8 $\%$ & $2.0 \cdot 10^{-8}$ & $1.5 \cdot 10^{-8}$&$1.1 \cdot 10^{-3}$\\
            \hline
           1500  & 98.6 $\%$ & $2.2 \cdot 10^{-9}$ & $1.6 \cdot 10^{-9}$&$6.7 \cdot 10^{-4}$\\
           \hline
           3000  & 99.2 $\%$ & $8.8 \cdot 10^{-11}$ & $6.5 \cdot 10^{-11}$&$3.3 \cdot 10^{-4}$\\
           \hline
        \hline
        \multicolumn{5}{|c|}{$\varepsilon=(3n)^{-2}$}\\
        \hline
           size $n$ & sparsity & RHS of \eqref{eq:sufficient1} & RHS of \eqref{eq:sufficient3} & $\varepsilon$  \\
            \hline
           300   &  89.2 $\%$ & $1.4 \cdot 10^{-7}$ & $1.0 \cdot 10^{-7}$ & $1.1 \cdot 10^{-5}$\\
            \hline
           900  & 95.6 $\%$ & $1.5 \cdot 10^{-8}$ & $1.1 \cdot 10^{-8}$&$1.2 \cdot 10^{-6}$\\
            \hline
           1500  & 97.1 $\%$ & $3.1 \cdot 10^{-9}$ & $2.3 \cdot 10^{-9}$&$4.4 \cdot 10^{-7}$\\
           \hline
           3000  & 98.4 $\%$ & $1.0 \cdot 10^{-10}$ & $7.7 \cdot 10^{-11}$&$1.1 \cdot 10^{-7}$\\
           \hline
           \hline
        \multicolumn{4}{|c|}{$\varepsilon=(3n)^{-3}$}\\
        \hline
           size $n$  & sparsity & RHS of \eqref{eq:sufficient1} & RHS of \eqref{eq:sufficient3} & $\varepsilon$  \\
            \hline
           300   & 84.3 $\%$ & $\mathbf{5.7 \cdot 10^{-8}}$ & $\mathbf{4.1 \cdot 10^{-8}}$&$3.7 \cdot 10^{-8}$\\
            \hline
          900  & 93.4 $\%$ & $\mathbf{1.1 \cdot 10^{-8}}$ & $\mathbf{8.1 \cdot 10^{-9}}$&$1.4 \cdot 10^{-9}$\\
            \hline
           1500  & 95.7 $\%$ & $\mathbf{1.5 \cdot 10^{-9}}$ & $\mathbf{1.1 \cdot 10^{-9}}$&$3.0 \cdot 10^{-10}$\\
           \hline
           3000  & 97.6 $\%$ & $\mathbf{5.9 \cdot 10^{-11}}$ & $\mathbf{4.4 \cdot 10^{-11}}$&$3.7 \cdot 10^{-11}$\\
           \hline
           \hline
        \multicolumn{4}{|c|}{$\varepsilon=(3n)^{-4}$}\\
        \hline
           size $n$ & sparsity & RHS of \eqref{eq:sufficient1} & RHS of \eqref{eq:sufficient3} & $\varepsilon$ \\
            \hline
           300   & 79.3 $\%$ & $\mathbf{4.0 \cdot 10^{-8}}$ & $\mathbf{2.8 \cdot 10^{-8}}$&$1.2 \cdot 10^{-10}$\\
            \hline
           900  & 91.3 $\%$ & $\mathbf{1.0 \cdot 10^{-8}}$ & $\mathbf{7.4 \cdot 10^{-9}}$&$1.5 \cdot 10^{-12}$\\
            \hline
           1500  & 94.3 $\%$ & $\mathbf{1.4 \cdot 10^{-9}}$ & $\mathbf{1.1 \cdot 10^{-9}}$&$2.0 \cdot 10^{-13}$\\
           \hline
           3000  & 96.8 $\%$ & $\mathbf{3.3 \cdot 10^{-11}}$ & $\mathbf{2.5 \cdot 10^{-11}}$&$1.2 \cdot 10^{-14}$\\
           \hline
        \end{tabular}
        \caption{\emph{For each value of $n$, one realization of the matrix $D$ is randomly generated, and $A,B,t$ are constructed as in Table \ref{tab1}. We then test, for various choice of $\varepsilon$ depending a priori on the size $3n$, what is the sparsity of $B_0$ (fraction of zero entries) and whether \eqref{eq:sufficient1} and \eqref{eq:sufficient3} are satisfied (the LHS, $\varepsilon$, is explicitly computed in the last column). When a test for exact recovery is satisfied, the corresponding values of the RHS are in bold font. }}
        \label{tab2}
    \end{table}

  {  \section{Numerical experiment on real-world data: Improving the efficiency of graph-based portfolio selection} \label{sec:realworld}
  
As discussed in the Introduction, the benefits of working with the complement graph for large, dense networks is well exemplified by applications that occur in a variety of contexts. In the present section, we test our results in  particular within one of these motivating applications, in the context of portfolio selection \cite{ANV,PDMA}.

 In our first experiment, we wish to compute the Katz centrality of a (possibly) very dense unweighted undirected graph of size $n=995$ nodes. We note that, while a network of $\approx 10^3$ nodes is not extremely large, optimizing the  complexity of computing centrality measures is still important, because for example one may want to train the method on old data before applying it to current data \cite{ANV}, and this would require that a graph centrality, say Katz, must be computed very many times within the overall algorithm. As described in \cite{ANV,PDMA}, one can form a correlation matrix of stock returns and subsequently threshold its entries based on a chosen threshold parameter. We will describe the process of constructing these matrices only in brief. For a more in-depth description of these matrices and of the underlying portfolio optimization methodology as a whole, we refer the reader to \cite{ANV,PDMA}. We form the correlation matrix $C$ of the historical returns, for the past $10$ years, of $995$ individual stocks within the S\&P1000 Dow Jones market index (the reason to select $995$ stocks, and not $1000$, is that we exclude those stocks that were not traded for the whole time period that we analyzed). Then, given a threshold parameter $\eta \in \left(-1,1\right)$, we form the adjacency matrix $A$ defined entry-wise as
\[A_{ij} = \begin{cases}
    1 & \mathrm{if} \ C_{ij} > \eta;\\
    0 & \mathrm{otherwise}.
\end{cases}\]
Having formed $A$ based on $\eta$, we then investigate the time taken to compute the Katz centrality for the resulting graph associated to $A$ and $B:= ee^T - A$ with a fixed Katz parameter taken to be $\alpha := 0.9/\rho(A)$. In both cases, we first form $A$ and $B$ and measure via MATLAB's \texttt{timeit} function the time taken to compute $(I-\alpha A)^{-1} e$ and $(I+\alpha B)^{-1} e$ using the backslash function. To ensure a fair comparison, we measure the time taken when $A$ and $B$ are stored as full and sparse matrices and select the minimum of the two times for each of the two matrices. Finally, we repeat this computation 100 times and plot the average time taken in Figure~\ref{fig:100runssp1000}. Different choices of $\eta$ lead to very different sparsities in the constructed network. For the data used in our experiment, the adjacency matrix $A$ ranges from very dense when $-1 < \eta \lesssim 0.1$ to very sparse when $0.55 \lesssim \eta < 1$. The figure clearly shows that, if $A$ is very dense, say $90\%$ or more of potential edges are present, then it is much more convenient to compute Katz centrality by working on the complement graph and using the results of this paper. Unsurprisingly, the opposite situation happens when $A$ is very sparse, thus suggesting that an optimized adaptive algorithm for portfolio optimization via unweighted graphs could test the network sparsity after selecting $\eta$, and switch accordingly between the graph and its complement. 
 Note that, by Theorem \ref{thm:trick1}, in this experiment all computations are exact up to numerical approximation due to the floating point implementation.

In our second experiment, we test the thresholding technique in the same spirit as in Table \ref{tab1}, but using real-world data. We produce a weighted graph from the same S\&P 1000 data as for the previous experiment. However, this time, we fix the parameter $\eta$ to $0$, and form the the following adjacency matrix $A$:
\[A_{ij} = \begin{cases}
    C_{ij} & \mathrm{if } \ C_{ij} > 0; \\ 0 & \mathrm{otherwise}.
\end{cases}\]
Here, $C$ is again the correlation matrix originating from the data and, as before, $A$ is very dense. Then, we form the adjacency matrix of the complement graph $B:= ee^T - A$. Due to the use of weights, it turns out that $B$ is itself very dense too, not allowing us to efficiently compute Katz centrality. Hence, we form the matrix $B_0$ as described in Subsection \ref{sec:weightedgraphs}, by setting to $0$ all entries in $B$ that are less than a tolerance $\varepsilon$. We then compute Katz centrality for $B_0$ and negative Katz parameter $-\alpha=-0.9/\rho(A)$, and compare it with that for $A$ and Katz parameter $\alpha$. In particular, we give the resulting Kendall's $\tau$ correlation coefficient \cite{kendall} for a variety of threshold values $\varepsilon$ in Table~\ref{tab:weighted}. It is worth noting that, for this particular graph, rather large values of $\varepsilon$ are required to significantly sparsify $B_0$, and thus to achieve an advantageous computational complexity. In this sense, the data that we have selected for the experiment lead to a graph that seems  extremely challenging for the thresholding technique proposed in Subsection \ref{sec:weightedgraphs}. Indeed, large values of $\varepsilon$ are far from satisfying \eqref{eq:sufficient1}, and in particular we cannot, a priori, be sure that the rankings computed from the corresponding $B_0$ will bear any resemblance to that arising from Katz centrality on $A$. However, it can be seen from Table \ref{tab:weighted} that in practice these ``approximate Katz rankings" remain remarkably well correlated with the exact one, even for very large values of the threshold $\varepsilon$. We remark that, in the context of the experiment and given the empirical evidence  of a high correlation even after extreme choices of the threshold $\varepsilon$, one may indeed be tempted to prefer efficiency over accuracy because, for the purposes of portfolio optimization, it suffices to  approximately rank the stocks \cite{ANV,PDMA}.

\begin{figure}
    \centering
    \includegraphics[width=\textwidth]{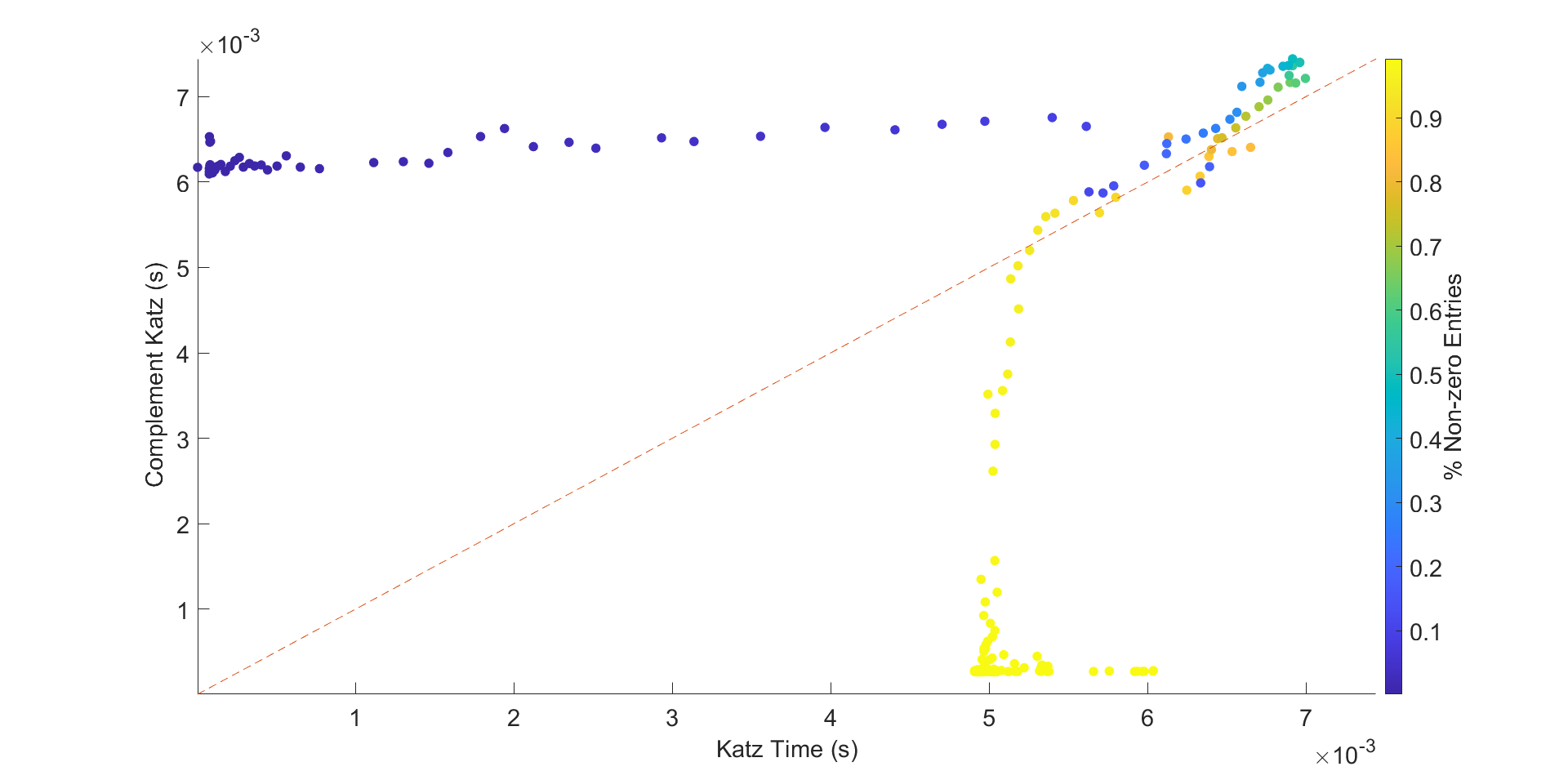}
    \caption{\emph{The average time to compute $(I-\alpha A)^{-1}e$ and $(I+\alpha B)^{-1}e$ for $\eta \in (-1,1)$, $\alpha = 0.9/\rho(A)$, for an unweighted financial graph representing historical correlations of S\&P1000 Dow Jones stocks. The range of times in the figure is 0-7 milliseconds.}}
    \label{fig:100runssp1000}
\end{figure}

\begin{table}[]
    \centering
    \begin{tabular}{|c|c|c|}
     \hline    Threshold, $\varepsilon$& Density of $B_0$ & Kendall's $\tau$  \\
         \hline 
 0 & 99.8995\% & 1\\
0.1 & 99.8839\% & 0.99986 \\ 
0.2 & 99.5207\% & 0.99643 \\ 
0.3 & 98.5381\% & 0.98928 \\ 
0.4 & 94.2217\% & 0.97672 \\ 
0.5 & 82.9872 \%& 0.96584 \\ 
0.6 & 64.5947\% & 0.95472 \\ 
0.7 & 42.2416\% & 0.92741 \\ 
0.8 & 21.2077\% & 0.88153 \\ 
0.9 & 7.6362\% & 0.79029 \\ 
\hline 
    \end{tabular}
    \caption{\emph{Table displaying Kendall's $\tau$ correlation coefficient between the rankings produced by Katz on a weighted graph of adjacency matrix $A$ (representing historical correlations of S\&P1000 Dow Jones stocks) Katz, with negative parameter, on its thresholded complement $B_0$, where $B_0$ is obtained by $B=ee^T-A$ by setting to zero its entries less than $\varepsilon$. The table also reports the sparsity of $B_0$ for a range of threshold values $\varepsilon$. The Katz parameter was set to $\alpha = 0.9/\rho(A)$. Note that $\varepsilon=0$ corresponds to $B_0=B$, and therefore Kendall's $\tau$ is exactly $1$, as expected by Theorem \ref{thm:trick3}.}}
    \label{tab:weighted}
\end{table}

}

    \section{Conclusions}

    We have discussed efficient algorithms to compute Katz and eigenvector centralities on very dense networks. The proposed techniques are based on computing Katz centralities, with negative parameter, on the complement graph of the given one. For unweighted graphs, our approach always guarantees exact computation of Katz centrality  in an effiicent manner for very dense networks. For weighted graphs, enhancement by thresholding may be needed, but we provided theoretical sufficient conditions for exact recovery and argued that, even when these are not met, good approximate recovery is still possible. We reported some numerical experiments that support our claims.

    {\section{Acknowledgements}

We are grateful to two anonymous reviewers, whose remarks have helped us to improve the presentation.}


\end{document}